\documentclass[aps,showpacs,prl,twocolumn]{revtex4}
\usepackage{graphicx}
\newcommand{\be}{\begin{eqnarray}}
\newcommand{\ee}{\end{eqnarray}}
\newcommand{\chil}{\chi_{\rm l}}
\newcommand{\chinl}{\chi_{\rm nl}}
\newcommand{\chisg}{\chi_{\rm sg}}

\begin{document}
\title{
Effect of random fluctuations on 
quantum spin-glass transitions 
at zero temperature
}
\author{Kazutaka Takahashi}
\author{Yoshiki Matsuda}
\affiliation{Department of Physics, Tokyo Institute of Technology,  
 Tokyo 152--8551, Japan}
\date{\today}

\begin{abstract}
 We study the effects of random fluctuations on quantum phase transitions
 by the energy gap analysis.
 For the infinite-ranged spin-glass models with a transverse field, 
 we find that a strong sample-to-sample fluctuation effect leads to 
 broad distributions of the energy gap.
 As a result, 
 the linear, spin-glass, and nonlinear susceptibilities 
 behave differently from each other.
 The power-law tail of the distribution implies 
 a quantum Griffiths-like effect
 that could be observed in various random quantum systems.
 We also discuss the mechanisms of the phase transition 
 in terms of the energy gap 
 by comparing  the Sherrington-Kirkpatrick model 
 and random energy model, which demonstrate 
 the difference between the continuous and discontinuous phase transitions.
\end{abstract}
\maketitle

 The quantum phase transition 
 that could occur by changing parameters in the Hamiltonian 
 is characterized by the closing of the energy gap between 
 the ground and first excited states.
 The emergence of the gap is due to a quantum effect and 
 we can find various new types of transitions  
 that are absent in 
 classical thermodynamic systems \cite{Sachdev}.
 The situation becomes more complicated when we treat disordered systems.
 We need to theoretically take an average 
 over realizations of disordered parameters.
 We expect that the self-averaging property holds 
 in the thermodynamic limit and that the average model describes 
 a real system represented by a specific realization.
 However, at low temperatures, 
 the sample-to-sample fluctuations become important 
 and lead to phenomena that can never be seen in clean systems.
 Actually, the gap vanishing point strongly depends on the random sample 
 and we can hardly identify the phase transition point.
 In a disordered phase of the system, 
 it is also known that physical quantities are affected 
 by the presence of finite clusters of ordered state.
 This Griffiths singularity \cite{Griffiths, McCoy} 
 is widely known in random spin systems
 as a general mechanism that induces random fluctuation effects.
 A similar effect is known in disordered electron systems 
 as anomalously localized states in delocalized phase \cite{ALS},
 which shows the universality of the phenomenon 
 for disordered systems.

 The Griffiths singularity in the transverse-field Ising spin-glass
 model was found numerically by 
 Rieger and Young \cite{RY}, and Guo {\it et al} \cite{GBH}.
 They found that the nonlinear susceptibility at zero temperature is 
 divergent below a point within the quantum paramagnetic phase 
 in two- and three-dimensional systems.
 It was argued that magnetically ordered clusters 
 in a sample gives a small gap,
 which makes the susceptibility large.
 The same effect was also found by Fisher \cite{Fisher}
 for the same model in one dimension.
 Although this effect was called the quantum Griffiths singularity, 
 the formation of the ordered cluster is essentially 
 the same as that in a classical case. 
 It is expected to disappear at large dimensions 
 since clusters easily form at low dimensions.

 It is well known that quantum systems
 can be formulated using path integral formalism.
 Then, the quantum effect can be represented by 
 fluctuations in an extra dimension of imaginary time
 and the averaging over disorder induces 
 correlations between different times.
 This is a common feature that can be seen 
 in all quantum systems with disorder.
 In this letter, 
 using simple quantum spin-glass models, 
 we consider a Griffiths-like mechanism 
 that can persist even in infinite-ranged models
 without the Griffiths singularity.

 We mainly discuss the Sherrington-Kirkpatrick (SK) model \cite{SK}
 in a transverse field 
 as one of the simplest quantum spin-glass model.
 The Hamiltonian is written as 
\be
 \hat{H} = -\sum_{i<j}^{N}J_{ij}\sigma^z_i\sigma^z_j
 -\Gamma\sum_{i=1}^N\sigma^x_i, \label{GSK}
\ee
 where $\sigma_i^{x,z}$ are Pauli matrices at site $i$, 
 $J_{ij}$ are random interactions with an average 
 $[J_{ij}]=0$ and a variance $[J_{ij}^2]=J^2/N$, 
 $N$ is the site number, and $\Gamma$ is the transverse field.
 Several analyses showed that 
 this model at zero temperature has a second-order phase transition 
 between the spin-glass and quantum paramagnetic phases 
 at $\Gamma/J\sim 1.4-1.5$ \cite{YI, MH, AR, ADDR, Takahashi}.
 The SK model is the two-body interacting one, and 
 one can generalize it into $p$-body ones. 
 The model at $p\to\infty$ is known to be equivalent to 
 the random energy model (REM) \cite{Derrida}.
 Its quantum version including the transverse field
 can be solved exactly and 
 shows a discontinuous transition 
 at $\Gamma/J=\sqrt{\ln 2}\sim 0.83$ \cite{Goldschmidt}.
 We treat both the SK model (\ref{GSK}) and the REM 
 since it is interesting to know 
 the nature of the phase transition
 from the energy spectrum.
 We numerically diagonalize their Hamiltonian matrices 
 by the Lanczos method.
 The number of spins, $N$, is taken up to 16.
 The corresponding size of the matrix is given by $2^N$.
 The ensemble average is taken over more than 20000 samples.

\begin{figure}[htb]
\begin{center}
\includegraphics[width=0.65\columnwidth]{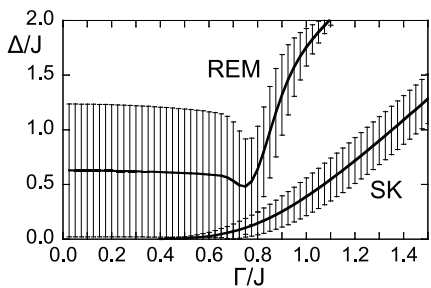}
\caption{Average energy gap $\Delta=[E_1-E_0]$
 between the ground and first excited states
 for the transverse SK model and REM ($N=16$).
 The variance of the energy gap $\pm\sqrt{[(E_1-E_0)^2]-\Delta^2}$ 
 is shown as the error bar. 
}
\label{gap}
\end{center}
\end{figure}

 In Fig.~\ref{gap},  
 we show the results of the average energy gap.
 In the SK model, the gap vanishing point is estimated as
 $\Gamma/J\sim 0.6$, which is much smaller than 
 the expected transition point.
 The gap vanishing point can be roughly estimated  
 using the perturbation theory as follows.
 When $J=0$, energy level is exactly given by 
 $E_n=-(N-2n)\Gamma$ with $n=0,1,\cdots,N$ and 
 each level has the degeneracy $N!/(N-n)!n!$. 
 When $J$ is included, these degeneracies are lifted 
 to form broad distributions of energy levels.
 The first excited state belongs to the sector $n=1$.
 In this sector, the matrix element of the $N\times N$ Hamiltonian 
 is given by $H_{ij}^{(1)}=-(N-2)\Gamma-J_{ij}$.
 The second term corresponds to the Hamiltonian of 
 the Gaussian random matrix theory, and 
 the energy levels are known 
 to form a semicircle distribution \cite{Mehta}.
 The width of the distribution is given by $4J$ and 
 the first excited energy level is given by $E_1\sim-(N-2)\Gamma-2J$.
 The unperturbed ground state with the sector $n=0$ 
 is given by $E_0\sim -N\Gamma$ and 
 the energy gap is obtained as $\Delta\sim 2(\Gamma-J)$.
 Thus, the degenerate point is roughly estimated as $\Gamma=J$ 
 which is modified by correlations between different sectors.
 Up to the second order in the perturbation theory, 
 the degenerate point is estimated as $\Gamma/J\sim 0.85$.
 Since the perturbation becomes worse at small transverse fields, 
 we cannot effectively predict the degenerate point.
 However, it can be definitely said that 
 the perturbative correction decreases 
 the value of the degenerate point and the point never 
 approaches the phase transition point at $\Gamma>J$.

 On the other hand, for the REM, we find a change in behavior 
 around the phase transition point.
 The average gap becomes minimum around the point and 
 the fluctuation also changes its behavior.
 This result is consistent with the analysis discussed in 
 Ref.~\onlinecite{JKKM}, where
 the transition point is supposed to be 
 equivalent to the point where the average gap becomes zero.

\begin{figure}[htb]
\begin{center}
\includegraphics[width=0.65\columnwidth]{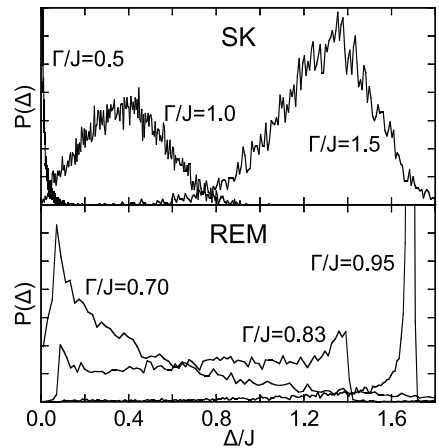}
\caption{Gap distribution function ($N=16$).}
\label{pg}
\end{center}
\end{figure}

 The gap fluctuation in Fig.~\ref{gap} can be clearly seen 
 from the gap distribution function in Fig.~\ref{pg}.
 For the SK model, we find a single peak distribution with a long tail, 
 which implies an important role of the fluctuation.
 In the case of the REM, we obtain more complicated distributions 
 that are dependent on $\Gamma$.
 For a large $\Gamma$, we observe a single peak 
 on the right-hand side of the distribution.
 When $\Gamma$ decreases, this peak decreases 
 and another peak at the left-hand side appears.
 The difference between the heights of the peaks 
 becomes smaller and changes its sign around the transition point.
 This behavior together with the result of the average gap implies
 a structural change between two different configurations
 at the transition point.

 In the classical spin-glass theory, 
 it is known that three types of susceptibilities,
 namely, linear $\chil$, spin-glass $\chisg$, and
 nonlinear $\chinl$, play important roles \cite{NL, FH}.
 When $\Gamma=0$,
 the linear susceptibility is represented by the spin-glass order
 parameter $q=[\langle \sigma_i^z\rangle^2]$ as
 $\chil=\beta (1-q)$, and 
 the nonlinear susceptibility can be expressed by
 the spin-glass one as
 $\chinl=\beta (\chisg-2\beta^2/3)$.
 The phase transition characterized by the divergence of $\chisg$
 can be found by observing the behavior of $\chinl$.

 These relations are changed in quantum systems 
 owing to the fluctuation effect.
 Using the spectral representation, we can write 
 the linear susceptibility as the sum of two contributions:
 $\chil=\chil^{(\rm c)}+\chil^{(\rm q)}$.
 The first term $\chil^{(\rm c)}=\beta (\chi-q)$ 
 corresponds to the classical part 
 and can be treated by the static approximation \cite{BM}.
 There is no classical counterpart for $\chil^{(\rm q)}$.
 At zero temperature, we have 
\be
 \chil^{(\rm q)} = \frac{2}{N}\sum_{n\ne 0}
 \frac{\left|\langle 0 |\sigma^z|n\rangle\right|^2}{E_n-E_0}, 
 \label{chil}
\ee
 where $|n\rangle$ is the eigenstate with the energy $E_n$, 
 $|0\rangle$ is the nondegenerate ground state, and
 $\sigma^z=\sum_{i=1}^N\sigma_i^z$.
 In the same way, 
 the quantum parts of $\chisg$ and $\chinl$ are 
\be
 & & \chisg^{(\rm q)}=
 \frac{4}{N}\left(
 \sum_{n\ne 0}\frac{\left|\langle 0 |\sigma^z|n\rangle\right|^2}{E_n-E_0}
 \right)^2,  \label{chisg}\\
 & & \chinl^{(\rm q)}
 = \frac{4}{N} \sum_{n,m\ne 0}
 \frac{|\langle 0|\sigma^z|n\rangle|^2}{E_n-E_0}
 \frac{|\langle 0|\sigma^z|m\rangle|^2}
 {(E_{m}-E_0)^2}
 \nonumber\\ & & 
 -\frac{4}{N} \sum_{n,n',n''\ne 0}
 \frac{ \langle 0|\sigma|n\rangle\langle n|\sigma|n'\rangle
 \langle n'|\sigma|n''\rangle\langle n''|\sigma|0\rangle}
 {(E_n-E_0)(E_{n'}-E_0)(E_{n''}-E_0)}.
 \label{chinl}
\ee
 The singularity of the susceptibility comes from the quantum part.
 It occurs when the energy gap 
 of the first excited level $\Delta$ reaches zero.
 Generally speaking, $\chil$, $\chisg$, and $\chinl$ depend on 
 $1/\Delta^{1,2,3}$, respectively, 
 and the energy gap distribution
 $P(\Delta)$ determines their critical behavior.
 When $\Delta$ is small, $P(\Delta)$ is assumed to have 
 the power-law form $P(\Delta)\sim \Delta^k$.
 Then, $\chil$,  $\chisg$, and $\chinl$
 diverge at $k\le 0$, $1$, and $2$, respectively.
 It should be stressed here that the point where $\chinl$ diverges 
 could be different from the point where $\chisg$ does.

\begin{figure}[htb]
\begin{center}
\includegraphics[width=0.65\columnwidth]{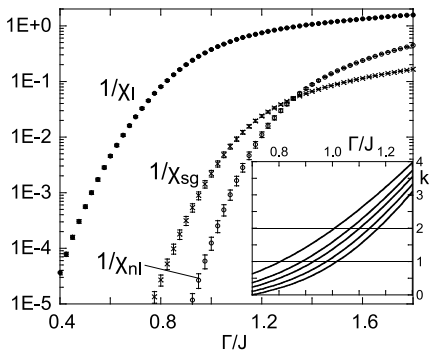}
\caption{Inverse of the average susceptibility (quantum part)
 for the SK model with $N=14$.
 Inset: The exponent $k$ of the energy gap distribution function
 ($N=8$,10,12,14, and 16 from above to below).
 The gap exponent is fitted by $P(\Delta)\sim \Delta^k$ near the origin.
}
\label{sus}
\end{center}
\end{figure}

 We plot the numerical result of the susceptibilities 
 and the exponent $k$ in the gap distribution function
 for the SK model in Fig.~\ref{sus}.
 The result implies that the three susceptibilities diverge 
 at different points, 
 which are roughly the same as the estimates from the exponent.
 Although the second term in Eq.~(\ref{chinl}) is neglected 
 in the calculation of $\chinl$, 
 we find in smaller systems that 
 the second term does not change the divergent point.
 We also find $J\chil\sim 2$ at a point where $\chisg$ 
 diverges, which is consistent with results of 
 the analysis discussed in Ref.~\onlinecite{ADDR}.

\begin{figure}[htb]
\begin{center}
\includegraphics[width=0.65\columnwidth]{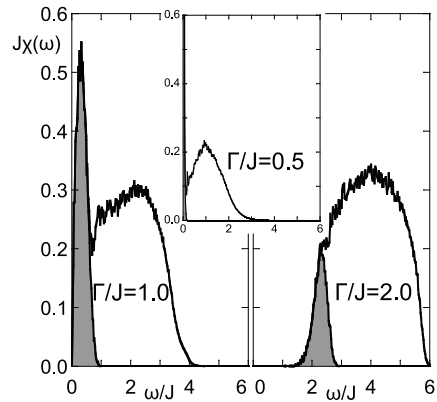}
\caption{Local correlation function $\chi(\omega)$ 
 for the SK model ($N=16$).
 The shaded areas are contributions of the first excited state.
}
\label{chi}
\end{center}
\end{figure}

 In order to make sure that 
 the gap between the ground and first excited states 
 determines the divergence of the susceptibilities, 
 we must examine the effect of higher-order excited states.
 For that purpose, we study the Fourier representations 
 of a real-time correlation function 
 including contributions from all states:
\be
 \chi(\omega) = 
 \sum_{n\ne 0}\delta(\omega-E_m+E_0)
 \left|\langle 0|\sigma_i^z|n\rangle\right|^2.
\ee 
 From this spectral function, we see 
 for the SK model that
 the main contribution comes from $N$ levels of the sector $n=1$ 
 if $\Gamma$ is sufficiently large. 
 This function was analytically calculated perturbatively 
 in ref.~\cite{TT} to yield a semicircle form.
 When $\Gamma$ decreases, 
 a sharp peak near the origin grows and 
 comes into contact with the vertical axis \cite{AR}.
 As shown in Fig.~\ref{chi},
 this peak is made from the first excited state only.
 We have the relation 
 $\chil^{(\rm q)}=\int d\omega\chi(\omega)/\omega$, 
 and the behavior of $\chi(\omega)$ around the origin 
 $\chi(\omega)\sim \omega^k$
 determines the divergence of $\chi$.
 Actually, the divergence of the quantity
 $\int d\omega\chi(\omega)/\omega^3$ 
 was used to find a phase transition \cite{ADDR}.
 Thus, we conclude that 
 the first excited state determines the critical behavior.

\begin{figure}[htb]
\begin{center}
\includegraphics[width=0.65\columnwidth]{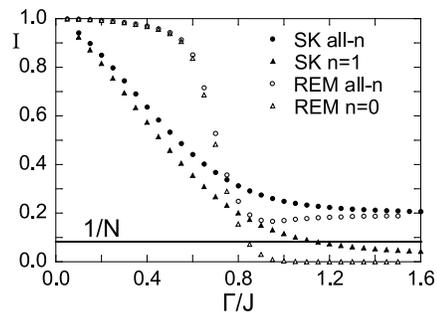}
\caption{Inverse participation ratio (N=12).
 The summation over $n$ in Eq.~(\ref{ipreq}) is restricted to 
 $n=0 (1)$  for a plot with the caption $n=0 (1)$.
 The error bars are smaller than the symbols.
}
\label{ipr}
\end{center}
\end{figure}

 The above analysis cannot be applied to the REM.
 In this model, $\chil$ and $\chinl$ do not diverge.
 They discontinuously jump at the
 transition point and stay constant in the spin-glass phase.
 The difference can be understood from 
 the analysis of the inverse participation ratio defined as 
\be
 I = \sum_{n}\left|\langle n|\sigma_i^z|0\rangle\right|^4.
 \label{ipreq}
\ee
 This quantity is estimated as the inverse of 
 the number of excited states included in $\sigma_i^z|0\rangle$.
 A similar quantity was used in disordered electron systems 
 to study the localization \cite{Wegner}.
 As shown in Fig.~\ref{ipr}, 
 when $\Gamma$ is small, 
 the difference between the SK model and the REM becomes transparent.
 The inverse participation ratio is close to unity in both cases  
 and the main contribution comes 
 from the first excited state for the SK model and 
 from the ground state for the REM.
 We also see from the distribution function of $I$ that 
 $I$ is always larger than $1/N$ which is consistent 
 with results of the analysis of $\chi(\omega)$ for the SK model.
 Thus, the gap analysis above is justified for the SK model.
 On the other hand, the contributions of the excited states 
 are very small and the classical analysis by static
 approximation is justified for the REM.

 We find that the linear, spin-glass, and nonlinear susceptibilities 
 diverge at different points, namely, $\Gamma_{\rm l}$, 
 $\Gamma_{\rm sg}$, and $\Gamma_{\rm nl}$ respectively.
 $\Gamma_{\rm sg}$ corresponds to the phase transition point. 
 Although our system size is not sufficiently large, 
 we roughly estimate that 
 the extrapolation to the $N\to\infty$ limit 
 gives $\Gamma_{\rm sg}/J\sim 1.4$, 
 which is consistent with the previous numerical results \cite{AR, ADDR}.
 In our numerical study, 
 $\Gamma_{\rm nl}/J\sim 1.6$ is somewhat larger than 
 $\Gamma_{\rm sg}$, 
 and $\Gamma_{\rm l}/J\sim 0.9$ is close to the point where 
 the average energy gap is zero.
 These behaviors can be verified by studying the distribution function
 of the susceptibility ${\rm P}_{\chi}(\chi)$.
 The power-law behavior ${\rm P}_{\chi}(\chi)\sim \chi^{-s-1}$
 with $s<1$ at large $\chi$ gives the divergence of 
 the average susceptibility.
 This distribution is roughly the same as 
 the power-law distribution of the gap 
 since $\chil\sim 1/\Delta$, for example.
 Our detailed study will be reported elsewhere.

 Although our result is very similar to that 
 obtained in finite-dimensional systems, 
 we cannot attribute the realization of  
 the small gap to the presence of ordered clusters.
 This is because no notion of the spatial dimension exists
 for the present infinite-ranged model.
 As we mentioned earlier, 
 the fluctuation in extra imaginary time direction 
 can induce the quantum Griffiths-like singularity  
 where an ordered cluster is randomly formed 
 in its direction, 
 which is similar to that obtained 
 in low-dimensional classical systems. 
 As in the finite-dimensional case \cite{Sachdev}, 
 by taking into account a rare event, 
 we find a power-law behavior of 
 the gap distribution function as
\be
 {\rm P}(\Delta)\sim\int d\ e^{-c\tau}\delta\left(
 \Delta-\Delta_0e^{-\frac{c}{k+1}\tau}\right)\sim \Delta^k,
 \label{qG}
\ee
 where $c$, $k$, and $\Delta_0$ are constants and 
 $\tau$ is the imaginary time.
 Instead of spatial directions 
 in the finite-dimensional case \cite{Sachdev}, 
 a small gap is caused by a large imaginary time $\tau$ 
 with an exponentially small probability.
 From the facts that 
 the randomness is absent in the imaginary time direction and 
 Eq.~(\ref{qG}) is relevant only for 
 the low-temperature limit of quantum systems,  
 it follows that the effect should be distinguished from 
 the Griffiths singularity. 
 Since the present effect is present 
 not only for the infinite-dimensional system 
 but also for finite-dimensional ones, 
 it is interesting to closely study 
 how both effects are distinguished in terms of physical quantities 
 in finite-dimensional systems.

 In conclusion, we have discussed quantum spin-glass transitions 
 in terms of the energy gap.
 In the SK model, our analysis implies that 
 three susceptibilities behave differently from each other  
 and diverge at different points, 
 which must be confirmed by further studies.
 The result can be characterized by 
 a power-law behavior of the distribution, 
 which we attribute to the quantum Griffiths-like effect.
 In contrast to the classical Griffiths singularity, 
 rare configurations of the quantum fluctuations 
 play a significant role in the phase transition.
 On the other hand, 
 the same mechanism cannot be applied to the REM.
 We have discussed conditions for this mechanism to be applied 
 by studying several quantities.
 It is important to state the conditions in a more general way,
 which will be part of our future investigation.

 We are grateful to 
 K. Takeda for helpful discussions and comments.
 Y.M. acknowledges the support from
 the Japan Society for the Promotion of Science.


\end{document}